# Pitfalls in VM Implementation on CHERI: Lessons from Porting CRuby

Hanhaotian Liu[a], Tetsuro Yamazaki[a], and Tomoharu Ugawa[a]

a    The University of Tokyo, Tokyo, Japan

**Abstract**    CHERI (Capability Hardware Enhanced RISC Instructions) is a novel hardware designed to address memory safety issues. By replacing traditional pointers with hardware capabilities, it enhances security in modern software systems. A Virtual Machine (VM) is one such system that can benefit from CHERI's protection, as it may contain latent memory vulnerabilities.

However, developing and porting VMs to CHERI is a non-trivial task. There are many subtle pitfalls from the assumptions on the undefined behaviors of the C language made based on conventional architectures. Those assumptions conflict with CHERI's stricter memory safety model, causing unexpected failures.

Although several prior works have discussed the process of porting VMs, they focus on the overall porting process instead of the pitfalls for VM implementation on CHERI. The guide for programming in CHERI exists, but it is for general programming, not addressing VM-specific issues.

We have ported CRuby to CHERI as a case study and surveyed previous works on porting VMs to CHERI. We categorized and discussed the issues found based on their causes.

In this paper, we illustrate the VM-specific pitfalls for each category. Most of the pitfalls arise from the undefined behaviors in the C language; in particular, implementation techniques and idioms of VMs often assume behaviors of traditional architectures that are invalid on CHERI. We also discuss workarounds for them and the impacts of those workarounds.

We verified the validity of the workarounds by applying them to our CRuby port and by surveying the codebases of prior case studies.

This work contributes to the body of knowledge on developing and porting VMs to CHERI and will help guide efforts toward constructing safer VMs.

# The Art, Science, and Engineering of Programming

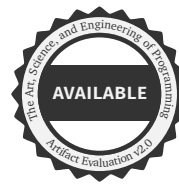
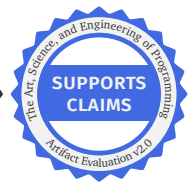

**Pitfalls in VM Implementation on CHERI: Lessons from Porting CRuby**

# 1 Introduction

Software security is important and has long been a topic for computer science research, particularly in preventing exploits from malicious parties [24, 26, 31]. Memory vulnerabilities remain a primary target for such attacks. Ensuring security is critical for all software, especially for Virtual Machines (VMs) of programming languages. They support high-level languages on top of them that constitute much of today's consumer services and applications. While VMs aim to provide safer execution environments for user programs, they themselves may contain bugs like any other complex software system [7, 23]. These flaws could undermine the security guarantees, including memory safety, of the higher-level language they support.

Maintaining security for VMs is challenging because of their characteristics. VMs are inherently complex with large code bases that bring broad attack surfaces. Thus, it is hard to identify and eliminate all potential vulnerabilities. On the other hand, the attacks are also constantly evolving, adding to the difficulty of ensuring security.

CHERI (Capability Hardware Enhanced RISC Instructions) [28, 29] is promising in preventing memory safety issues in VMs' implementations. It is an instruction set architecture (ISA) extension that introduces a capability-based security model and is implemented in RISC-V and ARM's Morello. By replacing traditional pointers represented by addresses with unforgeable, hardware-enforced capabilities, CHERI enables fine-grained memory protection, addressing spatial and temporal memory safety issues.

However, developing and porting VMs to CHERI is a non-trivial task, with many subtle pitfalls. Even for porting a lightweight interpreter, there are challenges, as shown in the porting of MicroPython [23]. The main difficulty arises from *undefined behavior* in the C language, which often behaves differently from VM developers' assumptions under CHERI. For example, conservative garbage collection (GC) relies on behaviors that are formally undefined but hold on conventional systems. Although the CHERI programming guide [30] warns of pitfalls and provides workarounds, its scope is general software, and it recognizes that porting language runtimes presents non-trivial challenges. Several previous works ported MicroPython [23], JavaScriptCore [16], and Rust [17], but they do not focus on pitfalls.

To explore the VM-specific pitfalls, we ported CRuby to CHERI as a case study as well as surveyed previous works. CRuby is a complex VM fully written in C. In our porting, we focus on the runtime system and the interpreter, excluding the JIT compiler. We also excluded the support for the foreign function interface (FFI). Although FFI is known as a source of security issues, we excluded it because we can isolate the executions of foreign functions by using CHERI's compartmentalization facilities as Chisnall et al. proposed for Java's Native Interface [11]. Currently, our porting of CRuby passes a total of 29,609/34,046 test cases included in the CRuby repository.

In this paper, we categorize issues that are specific to VMs on CHERI. For each category, we describe representative issues and discuss their causes along with possible workarounds. Although previous case studies have reported compiler-related and Rust-specific issues, this paper focuses on issues in runtime systems. Our research questions are as follows:

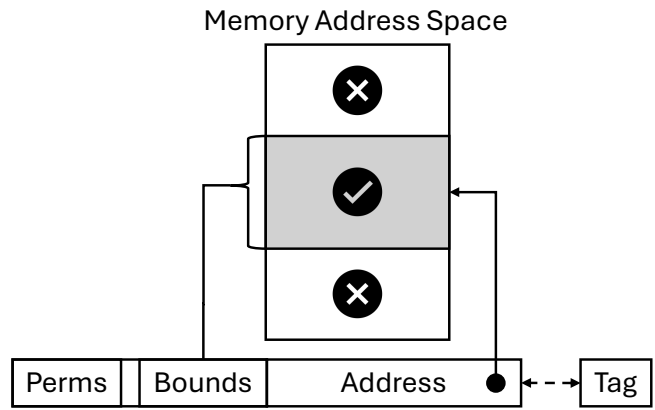

**Figure 1** Capability in CHERI.

**RQ1** What are the pitfalls specific to programming language VMs when developing and porting them to CHERI?

**RQ2** What are the possible workarounds for these pitfalls, and what are their impacts?

Since we focus on the porting of the VM itself, we do not further enhance VM security with CHERI's features, e.g., setting tight bounds on heap objects. It is left as a future work.

## 2 CHERI

We ported CRuby to the CHERI architecture. In this section, we describe the features of CHERI. CHERI provides two modes, the *purecap mode* and *hybrid mode*. Because our porting target is the purecap mode, we assume purecap mode throughout this paper.

### 2.1 CHERI Architecture

CHERI is an ISA extension that complements conventional hardware with architectural capabilities to achieve security features like fine-grained memory safety. It replaces pointers found in conventional architectures with capabilities and introduces tagged memory for tracking the validity of capabilities in the address space.

A capability is a classic computer science concept that represents a unforgeable token of authority [14]. In CHERI, a capability is implemented as a double-word fat pointer that contains an integer address and metadata, including bounds and permissions (as shown in Figure 1). The address part of the capability resembles that of a normal pointer found in conventional systems. The compressed bounds define the range that the capability is permitted to access, and the permissions control what operations are allowed on the authorized area in memory. Each capability is also associated with a 1-bit validity tag. The validity tag tracks whether the capability is valid to perform the earlier mentioned operations.

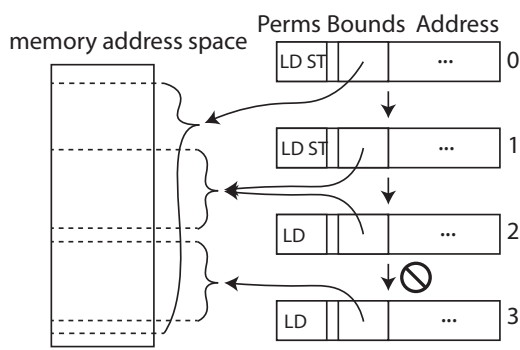

**Figure 2** Monotonicity of capability.

**Monotonicity** Monotonicity is one of the key security properties of capabilities. It ensures that the capability's permission and bounds cannot increase during derivation. The validity tag is preserved only when the capability is copied as-is, or when its permission and/or bounds are modified to be more restrictive.

Figure 2 illustrates an example of the monotonicity property. In the example, arrows indicate capability derivation. Capability 0 has the load and store permissions (LD/ST) to the memory region pointed to by it. Capability 1 is derived from capability 0. It has the same permissions as capability 0 but has bounds to a smaller region, so it is valid and can be constructed. Capability 2 is also valid because it is derived from capability 1 and has the same bounds but less permissions than capability 1. However, capability 3 cannot be constructed because it tries to have larger bounds than capability 2 from which it is derived.

**Sealing** Sealing is a property that makes a capability immutable and non-dereferenceable. Sealed capabilities should not be modified. The consequence for violating the sealing property is dependent on the semantics and the implementation of the CHERI architecture. It could result in clearing of tags or a hardware exception.

One form of capability sealing is the sealed entry capabilities. Sealed entry capabilities are capabilities to program code containing execution permission that are sealed. They are used to strengthen control-flow robustness [1] by preventing the undesired manipulation and use of those capabilities. CHERI-RISC-V creates sealed entry capabilities whenever it stores the value in program counter to a link register that is used for storing return addresses. Function pointers are also sealed entry capabilities.

## 2.2 CHERI C/C++ Runtime Environment

The CHERI team has developed a full software stack on top of the hardware implementations for supporting software development in the C/C++ languages on the CHERI platform [9]. It includes a capability-aware CheriBSD [10] operating system based on FreeBSD and a Clang/LLVM toolchain [8] that implements the CHERI C/C++ programming languages.

**Listing 1** Binary operation on (u)intptr_t variables.

```
uintptr_t
binop(uintptr_t lhs, uintptr_t rhs)
{
    uintptr_t res = lhs + rhs;
    return res;
}
```

CHERI C/C++ [30] are dialects of the widely used C/C++ languages. They provide spatial and temporal memory safety as well as extensions for accessing the capability features, including cheri_address_set for manipulating the address of a capability. Also, the address-of operator, "&", gives a narrow capability that points to the operand.

**(u)intptr_t and ptraddr_t Types** The (u)intptr_t integer types are for storing pointers in C/C++. They are also implemented as capabilities, so they are twice the size as on conventional architectures (128 bits). To store only the address part of the capability, CHERI C/C++ provides the new ptraddr_t type.

**Metadata Inheritance in Binary Operations** Since (u)intptr_t types are capabilities, they contain metadata. For binary operations, if only one operand is a (u)intptr_t variable, that (u)intptr_t variable's metadata will be inherited by the result. However, when both operands are (u)intptr_t variables, the metadata of the result is ambiguous. In such cases, it is designed that the left-hand-side operand's metadata will be inherited. An example of this is shown in Listing 1, where two uintptr_t variables are added together. The result res will inherit the metadata from the left-hand-side operand, lhs.

**Memory Management** In CHERI C/C++, malloc returns capabilities with bounds that only cover the allocated memory area. Revocation of a capability is the process of removing all copies and all derivatives of that capability from a program. The free function in CHERI C/C++ has been adapted to support revocation. When a capability is freed, the freed memory region is marked as being in quarantine and cannot be reallocated again until the periodic revocation is performed. After the revocation, all capabilities that point to the regions in quarantine are invalidated. This provides temporal memory safety against use-after-reallocation vulnerabilities.

Static memory allocations would also have bounds that only cover the memory area needed for the object. For example, if a 64-bit integer is statically allocated, the capability that represents its address will have bounds that only cover the 8 bytes it occupies.

## 3 Porting CRuby

In this work, we ported CRuby [2] to CHERI as a case study to investigate pitfalls in developing and porting VMs. CRuby, the reference implementation of the Ruby interpreter, has been used for executing various kinds of applications, web applications for example, since its first publication in 1995. As of 2025, the source code of CRuby is publicly available on GitHub,[1] and, roughly counting, consists of 470k lines of .c files and 130k lines of .h files.

Our porting was conducted as follows. We first compiled the source code using the CHERI C compiler and fixed errors whenever an error was reported during compilation. We also ran tests provided in the CRuby repository and fixed test failures. We ran all tests in directories basictest, bootstraptest, and test. Currently, our implementation passes all 894 test cases in basictest, 1,979 out of 1,987 test cases in bootstraptest, and 26,736 out of 31,165 test cases in test. We have not ported JIT compilers and the foreign function interface (FFI) for CRuby. The version of CRuby we ported was 3.4.1. For CHERI, we used the emulated CHERI-RISC-V. The total number of lines we modified in the CRuby codebase is 464 lines across 40 files. This corresponds to around 0.077 % of the whole codebase.

## 3.1 Design Choice

During the porting, besides the challenges that will be discussed in Section 4, ordinary porting tasks need to be done. For example, we defined macros included in the CRuby codebase for adapting to different platforms to correct values with respect to CHERI-RISC-V.

One design choice worth mentioning in our porting is that we defined `VALUE` as `uintptr_t`. In CRuby, `VALUE` is a data type denoting values in the Ruby language, though it is used across the CRuby codebase. It conveys different kinds of information, including pointers to objects, integers, and other data types. It is also used in various arithmetic and logical operations. In CRuby, `VALUE` is defined as an unsigned integer of some kind based on platforms, and during executions of the VM, it is frequently being cast to and from pointers. We define `VALUE` as `uintptr_t` since only `uintptr_t` is an unsigned integer type that can carry the pointer information in CHERI. This decision makes the size of `VALUE` different, increasing from 64 bits to 128 bits. Defining `VALUE` as `uintptr_t` is not unique compared to previous case studies. In the porting of MicroPython, the `mp_obj_t` type denoting a MicroPython object is represented by `uintptr_t` [23]. In the porting of JavaScriptCore, the `JSValue` type denoting values in the JavaScript language is also represented by capabilities [16].

We have not narrowed bounds of capabilities stored in `VALUE`s down to single object in our implementation. Instead, each `VALUE` has bounds fit to the heap page containing the object pointed to by that capability. In CRuby, a heap consists of one or more discontinuous heap pages and those memory regions are malloc'ed and free'ed in page units. The malloc function provided in standard CHERI C library assigns bounds to capabilities fit to the malloc'ed region. Thus, in current implementation, every `VALUE` has permission to access anywhere in the same heap page. Enhancing VM security by setting tight bounds on each object is left as a future work.

---

[1] https://github.com/ruby/ruby (accessed: 2025-07-17).

■ **Table 1** Categories of issues.

| Category | CRuby | JavaScriptCore | MicroPython | Rust |
|---|---|---|---|---|
| Invalid Derived Pointer | ● | - | - | - |
| Dereferencing Ambiguous Pointers | ● | ● | - | - |
| In-Place Reallocation | - | - | ● | - |
| Using Padding Bits of Integer Types | ● | ● | - | - |
| Modifying Temporary Capabilities Introduced by the Compiler | ● | ● | ● | - |
| Pointer Arithmetic on Non-Capability Type | ● | ● | ● | ● |

## 3.2 Failed Test Cases

There were in total 4,437 failed test cases. Of these, 2,931 failures originated from a single test program, which is a test for abstract syntax trees. This program crashed the test harness so that all the test cases included in the test program could not complete. Another major cause of failures was the absence of the "libyaml" library. 818 test cases depended on the "Psych" module, which requires it.

We have not yet identified the cause of the remaining 688 failed cases. They may contain unaddressed pitfalls in our porting process.

## 4 Pitfalls Across Various Ports to CHERI

During the porting of CRuby, we found several issues related to CHERI's capability model and its implementation. We categorized these issues, together with those reported in previous case studies, into several categories based on their causes, as shown in Table 1. We surveyed three case studies:

- JavaScriptCore [16]: porting of a production level JavaScript VM, with a JIT compiler, to CHERI.
- MicroPython [23]: porting of a small Python interpreter for embedded systems to CHERI.
- Rust [17]: porting of a Rust compiler, runtime, and libraries to CHERI.

Although previous case studies have reported compiler-related and Rust-specific issues, this paper focuses on issues in runtime systems. A circle in the table indicates that issues in the category were found in our porting or reported in previous case studies. This categorization allows us to understand the general nature of these issues and to derive general lessons for constructing and porting VMs to CHERI. While previous case studies reported issues in some categories, we also found issues in those categories that have not been reported.

Most of the issues are caused by assumptions about the undefined behaviors of the C language that CRuby makes. The C standard has a lot of undefined behaviors [22]. However, most C programs do not conform strictly to the C standard [25]. In particular, VMs often make assumptions about the undefined behaviors to realize language

**Pitfalls in VM Implementation on CHERI: Lessons from Porting CRuby**

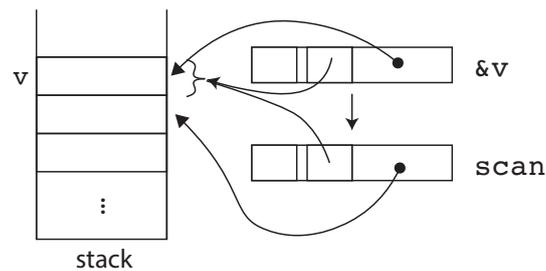

**■ Figure 3** Deriving stack top pointer.

**■ Listing 2** Pseudo code for stack scanning in CRuby.

```
1  void set_stack_end(VALUE **stack_end_p) {
2      VALUE top;
3      *stack_end_p = ⊤
4  }
5
6  void stack_scan() {
7      VALUE *scan = NULL;
8      set_stack_end(&scan);
9      while (...) {
10         VALUE v = *scan;
11         gc_mark(v);
12         scan++;
13     }
14 }
```

features efficiently. As a result, the software loses portability and fails to work on CHERI.

In the rest of this section, we discuss the categories in detail, showing some concrete examples of the issues in the categories. We also discuss possible workarounds and their impact on performance and security.

### 4.1 Invalid Derived Pointer

VMs often derive a pointer from another that is created for a different purpose. For example, CRuby takes the address of a local variable on the stack and derives a pointer for stack scanning. This works on conventional architectures, where all pointers are memory addresses. However, in CHERI, such practices are not valid because pointers are capabilities with bounds.

**Issue** CRuby uses a conservative GC. During root scanning, the conservative GC needs to determine the range of the stack. It obtains the stack top address from that of a local variable, as shown in Listing 2. Because the frame of set_stack_end is on the top of the stack when it is called, the address of the local variable, &top, is the stack top address, as illustrated in Figure 3. The stack scanning routine, stack_scan, iterates from the stack top to the bottom using the pointer, scan, which is initialized with the stack top address. During this iteration, it increments scan, thereby deriving a pointer.

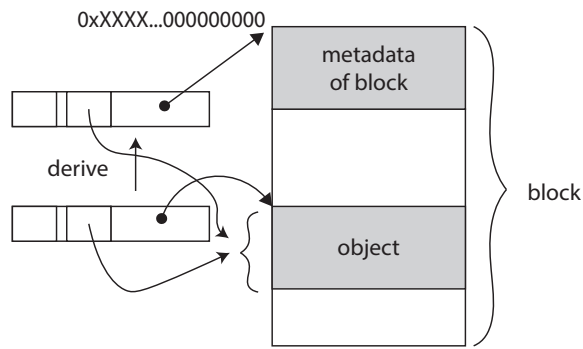

**Figure 4** Deriving pointer in block-structured heap.

This code is invalid in CHERI because the "&" operator returns a capability whose bounds cover only the local variable top. Consequently, the derived pointer scan also inherits bounds that are limited to top. When the stack scanning loop dereferences scan in the second iteration, it causes a capability bounds fault because scan is out of bounds.

**Possible Issue** This kind of issue could also arise in memory managers of VMs. For example, in a block-structured heap [15], a memory manager sometimes stores metadata at the beginning of a memory block. To access the metadata of the block, it derives a pointer to the beginning of the block, which is aligned to a $2^n$-boundary, from a pointer to an object in the block as shown in Figure 4. If a pointer to the object had bounds that cover only the object, the derived pointer to the metadata is invalid. Although CRuby also employs a block-structured heap, we did not narrow pointer bounds to individual objects in our porting. Instead, we maintain bounds that cover the entire block. Therefore, this issue did not arise.

**Workaround and Its Impact** Our solution in porting CRuby is to keep a *super capability* with sufficiently wide bounds to cover the entire memory region that may be accessed in the future, such as the entire stack. When the VM needs to create a pointer that operates on a region covered by the super capability, the new pointer can be safely derived from it. In the case of stack scanning, the CHERI C compiler already maintains the stack pointer to cover the entire stack. Thus, we create scan in Listing 2 from the stack pointer.

One potential drawback of this workaround, in general, concerns the security of software running on CHERI. Circulating a capability that covers a wide memory region could introduce vulnerabilities. To mitigate this, we could compartmentalize the routine that requires the super capability, as Chisnall et al. proposed for Java's Native Interface [11], thereby limiting its impact. However, compartmentalization incurs its own overhead, leading to a trade-off between security and performance.

**Pitfalls in VM Implementation on CHERI: Lessons from Porting CRuby**

**Lesson**   When deriving a pointer from another one, we have to make sure that the original pointer has enough bounds and permissions. If not, we have to keep a super capability that has enough bounds and permissions.

## 4.2 Dereferencing Ambiguous Pointers

In the C standard, when we convert an integer to a pointer type, the result is implementation-defined (Section 6.3.2.3 in C standard [19]). The converted pointer may be invalid, and dereferencing it causes undefined behaviors. In CHERI, converting a non-pointer value to a pointer does not yield a valid capability, so dereferencing it causes a capability tag fault. However, conservative GC assumes that all values on the stack that look like pointers are potential pointers and attempts to dereference them. Such pointer-like values are called *ambiguous pointers*.

**Issue**   Conservative GC [4] is an implementation technique for GC that does not require precise information about the locations of pointers. It scans the stack for values that look like pointers and handles them as roots.

Listing 2 shows the pseudo code of stack scanning in CRuby. In `gc_mark`, it tests whether every value aligned to the pointer size, v, looks like a pointer and dereferences it if so. The test includes checking whether v has the bit pattern representing the starting address of an object in the heap. However, if there is an integer that has exactly the same bit pattern as the address to an object, the GC will convert this integer to a pointer and attempts to dereference it. In CHERI, such a converted pointer does not have the validity tag set. Therefore the attempt of the dereference fails.

The same issue has been reported in the case study of porting JavaScriptCore (Section 4.1.2 in [16]). JavaScriptCore also scans the stack conservatively, and without fix, it caused tag violation during GC.

**Workaround and Its Impact**   In CHERI, we can tell precisely if a value in a memory location is a pointer or not by referring to the associated validity tag. The workaround for this issue is to test whether the validity tag of the ambiguous pointer being dereferenced is set or not, instead of the bit-pattern based test. This workaround may rather improve performance because it allows us to avoid marking dead objects that would be reachable only when we dereference pointer-like integers. Such pointer-like integers do exist; during the execution of the basictest test suite, we observed that 350 pointer-like integers passed the check by CRuby's GC to filter out non-pointers before marking. Wang et al. [27] also reported that, in their copying GC for Ruby, they needed to pin objects pointed to by stack pointers to avoid incorrectly updating pointer-like integers. Precise scanning using validity tags opens up opportunities to implement a copying GC without pinning, as well.

**Lesson**   Conservative GC in CHERI must test the associated validity tag before dereferencing an ambiguous pointer.

■ **Listing 3** Pseudo code of a part of parser in MicroPython.

```
 1  void* parser_alloc(parser_t parser, size_t num_bytes) {
 2      chunk_t chunk = parser->cur_chunk;
 3      if (chunk->used + num_bytes > chunk->alloc) {
 4          size_t new_size = sizeof(chunk_t) + chunk->alloc + num_bytes;
 5          chunk_t new_data = m_renew_maybe(chunk, new_size);
 6          if (new_data != NULL) {
 7  #ifdef __CHERI_PURE_CAPABILITY__
 8              parser->cur_chunk = chunk = new_data;
 9  #endif
10              chunk->alloc += num_bytes;
11          } else
12              chunk = NULL;
13      }
14      if (chunk == NULL) {
15          parser->cur_chunk = chunk = m_new(...);
16          initialize chunk;
17      }
18      byte* ret = ((byte*) chunk) + sizeof(chunk_t) + chunk->used;
19      chunk_used += num_bytes;
20      return ret;
21  }
```

### 4.3 In-Place Reallocation

In the C standard, the realloc function deallocates the old object pointed to by the pointer passed as the arguments (Section 7.20.3.4 in C standard [19]). Therefore, operations on that pointer after returning from realloc are undefined. This is the case even when the returned pointer from realloc points to the same address by chance. In fact, realloc may return a capability with different bounds from the one that is passed to realloc in CHERI. The relevant instruction is described in the CHERI C/C++ Programming Guide (Section 4.3.2 in [30]).

This is the case for custom allocators. VMs often make custom allocators that tell whether in-place reallocation succeeds or not, or even guarantee in-place reallocation in special use cases. The client code of such a custom allocator may keep using the original pointer when the in-place reallocation succeeds. This does not work in CHERI if the custom allocator assigns tight bounds to the pointers that returns because, in cases of expanding reallocation, the bounds of the old pointer do not cover the expanded area.

**Issue** This case is described in the porting of MicroPython (Section 4 in [23]). In its parsing and string interning processes, it allocates a chunk of memory from GC and divide it for real objects. Listing 3 shows the pseudo code of the parser, where __CHERI_PURE_CAPABILITY__ is the fix for this issue. When it runs out of the current chunk, it attempts to reallocate in-place for a chunk using m_renew_maybe as shown in Figure 5. If the attempt of in-place reallocation succeeds, m_renew_maybe returns the new pointer, which is identical to the original pointer in the original MicroPython. Thus, the original MicroPython discards the returned pointer, new_data, when it is not NULL and keeps using the original pointer, chunk. However, in the CHERI porting, the returned pointer is no longer identical to the original one; it has wider bounds. Therefore, writing to the expanded area using the pointer that parser_alloc returns results in a capability bounds fault.

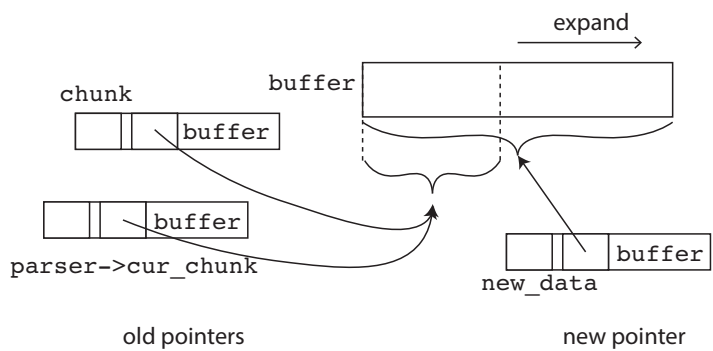

**Figure 5** In-place reallocation in MicroPython.

**Workaround and Its Impact**  As shown in Listing 3, the CHERI porting of MicroPython reassigns the returned pointer to all locations that hold the old one. In this case, those locations are parser->cur_chunk and the local variable chunk. Note that all pointers previously returned from parser_alloc do not need to be updated, even if they point to the middle of the current chunk, because expanded area will not be accessed through them.

In the general case, the old pointer may be stored in many locations, and sometimes it is difficult to enumerate the locations statically. In that case, we might need a special mechanism to keep track of those locations. Another possible workaround would be to access the memory area that may later be reallocated through indirection. Both involve substantial performance overhead.

The best workaround is not to rely on in-place reallocation, even with custom allocators. In theory, in-place reallocation is merely an optimization that can be applied at the allocator's discretion. Client code should not depend on it, even from the perspective of modularity. Although one could design an API that guarantees in-place reallocation while allowing client code to keep using the original pointers on traditional architectures, this is impossible on CHERI.

**Lesson**  When we develop VMs for CHERI, we should avoid depending on in-place reallocation.

### 4.4 Using Padding Bits of Integer Types

VMs often pack small values into integers, such as bitmaps or manually packed bit-fields computed through explicit bit operations. VMs also regard an array of smaller integers as an array of large integers to exploit data-parallelism. In such cases, programmers sometimes assume that all bits of the integer value are available. However, some bits may be *padding bits* (Section 6.2.6.2 in C standard [19]). The values of any padding bits are unspecified, and they could be tied to machine-specific constraints. In CHERI C, one of the integer types, (u)intptr_t, is implemented with capabilities, whose upper 64 bits hold metadata, such as bounds. The corresponding bits of (u)intptr_t are padding bits, and only the lower 64 bits of (u)intptr_t can be used to represent integer values.

Furthermore, the results of the shift operations, which are often used in these situations, are undefined if the shift amount is greater than or equal to the *width* of the type, which is the number of bits excluding the padding bits (Section 6.5.7 in C standard [19]). In CHERI C, we observed that left shift operations on (u)intptr_t values yields a random value if the shift amount is 64 or more.[2]

**Issue 1** CRuby uses bitmaps for marking live objects in GC. The bitmaps are declared as arrays of uintptr_t. To locate $i$-th bit in the bitmap, CRuby computes $index = \lfloor i/S \rfloor$ and $bit\text{-}offset = i \bmod S$, where $S$ is the number of bits in uintptr_t. By using *index* and *bit-offset*, it sets $i$-th bit in the following way:

   bitmap[*index*] |= 1 << *bit-offset*;

Here, $S$ is computed from sizeof(uintptr_t), which includes the padding bits. Thus, *bit-offset* can be greater than or equal to the width of uintptr_t, 64, and in such cases, left shifting 1 by *bit-offset* bits fails. Even if left-shifting successfully yielded a correct value, which is greater than or equal to $2^{64}$, the result of the bitwise-or operation is undefined. In CHERI, the upper half bits will not change.

**Issue 2** CRuby embeds shape_id, an identifier of *hidden class* [6], in the higher bits of an object header, which is of a capability type, VALUE. Therefore, the following code that installs shape_id does not work:

   #define SHAPE_FLAG_SHIFT  ((SIZEOF_VALUE * 8) - SHAPE_ID_NUM_BITS)
   RBASIC(obj)->flags |= ((VALUE)(shape_id) << SHAPE_FLAG_SHIFT);

Related issues are reported in the porting of JavaScriptCore [16].

**Issue 3** When CRuby counts the number of UTF-8 lead bytes in a string, it exploits data parallelism. Listing 4 shows the pseudo code for the function that counts it. This function counts the number of bytes that have either 0 at the most significant bit or 1 at the second most significant bit in location, s, which is of type uintptr_t. The caller of this function iterates over a string using uintptr_t pointer to process sizeof(uintptr_t) bytes at a time. However, half of the bits of uintptr_t are padding bits in CHERI. Thus, this function operates on only half of the bits as shown in Figure 6 and returns an incorrect count.

---

[2] If the compiler finds that the shift amount is a large constant, it eliminates the shift instruction, while keeping using the output register of the eliminated instruction as the result of shifting.

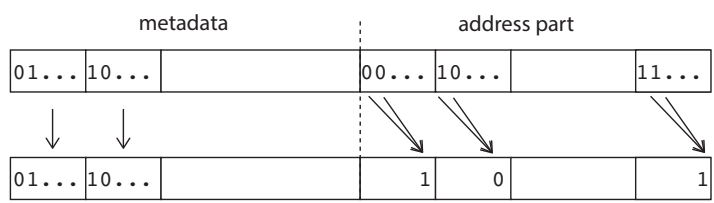

**Figure 6** Failure in parallel counting of UTF-8 leading bytes.

**Listing 4** Pseudo code for counting UTF-8 lead bytes.

```
1  count_utf8_lead_bytes_with_word(uintptr_t* s) {
2      uintptr_t d = *s;
3      d = (d >> 6) | (~d >> 7);
4      d &= NONASCII_MASK >> 7; // NONASCII_MASK is 0x808080....
5      return rb_popcount_intptr(d);
6  }
```

**Workaround and Its Impact** The solution is to use *exact-width integer types*, such as uint64_t, for those usages. Such types are specified to have designated widths and no padding bits (Section 7.18.1.1 in C standard [19]). This workaround will not introduce performance overhead.

**Lesson** We should use exact-width integer types for bitmaps and manually packed bit-fields computed through explicit bit operations.

### 4.5 Modifying Temporary Capabilities Introduced by the Compiler

Capabilities in CHERI can be *sealed* to prevent them from being modified as we described in paragraph 2.1. For example, return addresses on the stack are sealed. Any attempt to modify sealed capabilities will cause a *capability sealed* fault.

Modifications of capabilities can be introduced by the CHERI C compiler even if the programmer does not explicitly modify them. In CHERI C, a binary operation with one operand of type (u)intptr_t yields a (u)intptr_t value. Because (u)intptr_t in CHERI C is implemented with capabilities, the result inherits metadata of the operand of type (u)intptr_t. To achieve this, the CHERI C compiler generates code that copies the (u)intptr_t operand and replaces its address part with the result of the binary operation. If the (u)intptr_t operand is converted from a sealed capability, this binary operation causes a *capability sealed* fault.

**Issue 1** When CRuby generates a backtrace on an error, it collects return addresses using libunwind. The collected return addresses are sealed entry capabilities. CRuby performs operations on them to find the symbols of the functions that contain those return addresses. Listing 5 shows the pseudo code to find the symbol of function that contains the address trace_addr, and Figure 7 is its illustration. For every (function) symbol, it computes the symbol's address saddr. And then, it computes d, the distance

■ **Listing 5** Finding symbol for function containing trace_addr.

```
1  for (j = 0; j < symtab_count; j++) {
2      ElfW(Sym) *sym = &symtab[j];
3      uintptr_t saddr = (uintptr_t)sym->st_value + base_addr;
4      uintptr_t d = (uintptr_t)trace_addr - saddr;
5      if (d < (uintptr_t)sym->st_size) {
6          /* sym is the symbol of the function containing trace_addr */
7          /* store the information of sym in output buffer */
8          break;
9      }
10 }
```

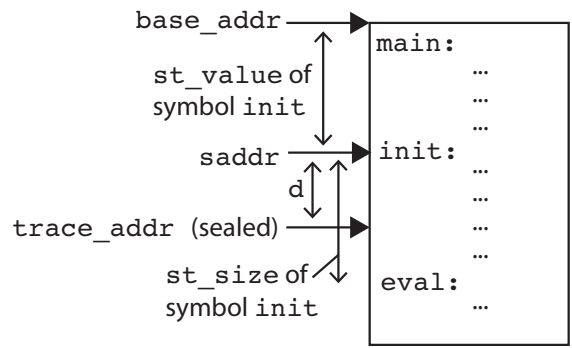

■ **Figure 7** Finding symbol for trace_addr (Illustration of Listing 5). The function is init in this example.

■ **Listing 6** Hash function.

```
1  st_numhash(st_data_t n)
2  {
3      enum {s1 = 11, s2 = 3};
4      return (st_index_t)((n>>s1|(n<<s2)) ^ (n>>s2));
5  }
```

between trace_addr and saddr. The latter computation attempts to modify the sealed entry capability trace_addr, and causes a *capability sealed* fault.

**Issue 2** CRuby supports a direct threaded code interpreter [3]. The instruction sequence of a threaded-code interpreter is not an array of instruction IDs, such as bytecode numbers. Instead, it consists of an array of addresses of runtime routines, each of which interprets a single bytecode instruction, followed by their operands. However, VMs sometimes need the ID of the current instruction for several reasons, such as debugging. Thus, CRuby creates a hash table to find the bytecodes' information during initialization. The keys of the hash table are the addresses of the runtime routines, which are sealed entry capabilities. Therefore, the hash function in Listing 6 attempts to modify a sealed entry capability, causing a *capability sealed* fault.

**Pitfalls in VM Implementation on CHERI: Lessons from Porting CRuby**

■ **Listing 7** Testing bits in VALUE.

```
1  int RB_IMMEDIATE_P(VALUE v) {
2      return (v & RUBY_IMMEDIATE_MASK) != 0; // RUBY_IMMEDIATE_MASK is 0x7
3  }
4
5  gc_mark(VALUE v) { // v may be a return address on the stack
6      if (RB_IMMEDIATE_P(v))
7          return;
8      // further tests if v looks like a pointer and marks v's referent
9  }
```

**Issue 3**   In CRuby, the VALUE type has tag bits in its lower three bits, which are all zero if the VALUE is a pointer. The RB_IMMEDIATE_P function in Listing 7 tests those bits and tells if the given value, v, is an immediate (not a pointer) or not. Because we define VALUE as uintptr_t, v is a capability. Thus, the bit-wise and operation in the test updates the address part of v to create a temporary capability that inherits metadata from v. The address part of this temporary capability is tested to determine if it is zero or not. Note that compilation with -O1 or higher optimization does not create this temporary capability.

The conservative GC calls this RB_IMMEDIATE_P function to test whether each value on the stack looks like a pointer or not. Thus, return addresses on the stack are also passed to RB_IMMEDIATE_P. However, return addresses are sealed. Therefore, the bit-wise and operation fails to update the address part of v. Note that the workaround for dereferencing an ambiguous pointer in Section 4.2 is not sufficient for this issue because return addresses are valid capabilities.

**Workaround and Its Impact**   We converted the (u)intptr_t operands to non-capability types such as uint64_t before the operation as below:

   return (((uint64_t) v) & RUBY_IMMEDIATE_MASK) != 0;

This workaround prevents the compiler from introducing a temporary capability. Although the results of the operations cannot be converted back to pointer, this workaround works because the results of the operations are only used as integers.

This workaround does not introduce any performance overhead. Rather, this workaround eliminates unnecessary instructions to create a temporary capability. This is because the compiled code first takes the address part of the operand and performs the operation on it as if it is an integer type. Then it updates the address part of the operand to create the temporary capability. However, this temporary capability is not used as a capability, but only its address part is used as an integer in these issues.

This workaround does not work for the case where the results need to be converted back to pointers. However, the operations should not be allowed in this case if their operands are sealed capabilities because these operations are attempts to create derived pointers from sealed capabilities.

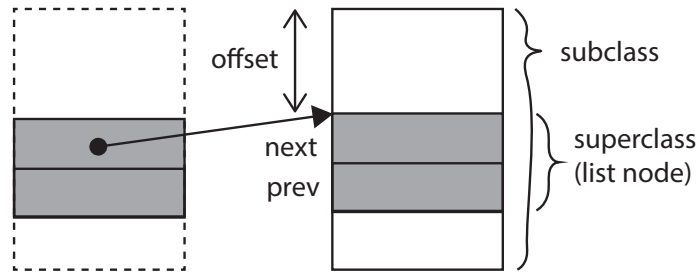

**Figure 8** Class inheritance and downcast in C.

Note that, in newer versions of CHERI,[3] the issues in this section would not cause a runtime exception because the semantics for modifying the address of sealed capabilities have been changed to invalidating the capability. However, this workaround is still valid to avoid producing unnecessary temporary capabilities.

**Related Issues** The CHERI C compiler reports a warning if both operands of a binary operation are capabilities, such as (u)intptr_t. This is because it is ambiguous that from which operand the result should inherit metadata. MicroPython converted (u)intptr_t to integer types to suppress these compiler warnings [23].

**Lesson** When performing operations on (u)intptr_t to obtain an integer result that is not necessarily (u)intptr_t, the (u)intptr_t operand should be cast to the result type before the operation.

### 4.6 Pointer Arithmetic on Non-Capability Type

Pointer arithmetic is common in VM implementations. Typically, a pointer is cast to an integer, and after computations on the integer, it is cast back to a pointer. In the C standard, the result of this operation is implementation-defined (Section 6.3.2.3 in C standard [19]). In CHERI, this works only when the integer type is (u)intptr_t and the original pointer has bounds that cover the new address. This is reasonable because (u)intptr_t is the only integer type that guarantees the recovery of the original pointer when it is cast back without modification. However, programmers sometimes use integer types defined for other purposes, such as size_t.

**Issue** Emulating class inheritance in C by embedding instances of superclasses in an instance of subclass is a common idiom. A typical use case is to embed a list node structure, which contains prev and next members, in a subclass instance as illustrated in Figure 8. The subclass instances are maintained as a list of such list nodes, using a common list-manipulation routine. When a node is retrieved from the list, it is *downcast* to the original subclass. This idiom is frequently observed in operating systems [21].

---

[3] The semantics were changed in version 9.

■ **Listing 8** Downcast.

```
1  static rb_thread_t *
2  thread_sched_waiting_thread(struct rb_thread_sched_waiting *w)
3  {
4      if (w) {
5          return (rb_thread_t *)((size_t)w - offsetof(rb_thread_t, sched.waiting_reason));
6      }
7      // …
8  }
```

The same idiom is used in the thread manager in CRuby. CRuby's thread structure rb_thread_t has a member sched.waiting_reason. This member behaves as a superclass of rb_thread_t. Listing 8 shows the pseudo code of downcasting to recover the pointer to rb_thread_t from the given w, the address of the member sched.waiting_reason. The macro offsetof yields the offset of this member from the top of the rb_thread_t. The problem is that this code casts w to size_t for pointer arithmetic. Thus, casting back to rb_thread_t yields an invalid capability.

**Issues in MicroPython and Rust**   Similar issues were also observed in the porting of MicroPython [23] and the Rust compiler [17]. MicroPython does not cast uintptr_t to the universal type mp_obj_t, but it first casts uintptr_t to mp_uint_t before casting to mp_obj_t. Because mp_uint_t is not a capability type in the MicroPython porting, the resulting mp_obj_t value cannot be used as a pointer even if the original uintptr_t is converted from a pointer.

In Rust, a library provides an inter-thread communication channel, which can send and receive integers. To send a pointer using the channel, the library casts pointers to integers before passing and cast back to pointers afterwards. In CHERI, the received pointer cannot be dereferenced.

**Workaround and Its Impact**   Our workaround for this issue is to change size_t to uintptr_t. This workaround introduces additional instructions to manipulate capabilities, which are necessary to produce a valid capability.

**Lesson**   In CHERI, (u)intptr_t should be used for pointer arithmetic. More generally, pointer arithmetic on integer types is not portable.

### 4.7 CHERI's Issues

This last category of issues is about the CHERI hardware and its software stack. CHERI is relatively new and under active research and development. As with any complex

■ **Table 2** Environment used for running benchmarks.

| Attribute | Value |
| --- | --- |
| CPU Model | 12th Gen Intel(R) Core(TM) i9-12900 |
| CPU Frequency | Fixed to 2.4 GHz, turned off Turbo Boost and dynamic frequency scaling |
| CPU Settings | Isolated one logical CPU and turned off Hyper-Threading |
| Operating System | Ubuntu 22.04 |
| clang Version | 15.0.0 |

system under development, it may contain problems. In this subsection, we report the CHERI's issues that we observed during porting CRuby.[4]

**mprotect** CRuby temporarily protects a page from any access using mprotect during GC. However, we observed that, once we remove permission from a page, all the capabilities on the page are invalidated after we recover the permission. In a newer version,[5] capabilities are maintained valid if we pass the PROT_CAP flag to mprotect when we recover the permission.

**makecontext** CRuby uses makecontext to implement coroutines. The makecontext function declared below takes a function func and an arbitrary number of additional arguments and makes a *context* holding func and the additional arguments.

```
void makecontext(ucontext_t *ucp, void (*func)(), int argc, ...)
```

The function func can later be called with the additional arguments using the created context. However, makecontext copies additional arguments to the context as uint64_t. Therefore, if some of the additional arguments are capabilities, they are invalidated and cause capability tag faults when func attempts to dereference them.

## 5 Performance Evaluation

To illustrate the performance overheads introduced by the workarounds in our ported Ruby, we compared the performance of original Ruby VM and our ported Ruby VM. Both were of version 3.4.1.

### 5.1 Benchmarks

We used 894 micro-benchmark programs included in the official Ruby GitHub repository.[6] These benchmarks measure the performance of various aspects of the Ruby VM, including method calls, arithmetic operations, class object operations, and application of common algorithms. We ran all benchmarks except for those in the gc

---

[4] We confirmed that these are issues of CHERI by personal communication in CHERI-CPU Slack.
[5] 25.03 release.
[6] https://github.com/ruby/ruby/tree/master/benchmark (accessed: 2025-11-07).

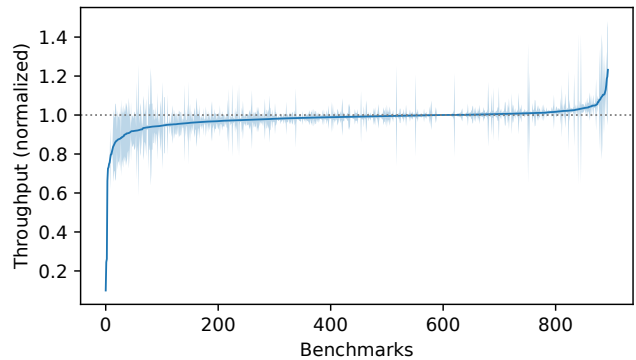

■ **Figure 9** Throughput ratio of the ported Ruby against the original Ruby with ±1$\sigma$. Higher is better.

subdirectory, which were excluded from the default setting. We also excluded 3 benchmarks, dump_all_file, vm_fiber_allocate, and vm_fiber_count, for which original Ruby VM reported errors.

## 5.2 Experimental Setup

The environment setting that we used to run the benchmarks are shown in Table 2. We run both VMs on an x86 architecture CPU.

To make the ported Ruby compilable and runnable on the x86 architecture, we modified it minimally and preserved the prior modifications for porting. The GC effects cannot be shown as pointer validity cannot be checked, so both versions run GC in a conservative fashion.

## 5.3 Evaluation

Figure 9 plots the throughput the ported Ruby VM normalized to the original Ruby VM for each benchmark. The higher is better. The x-axis is sorted by the mean throughput ratio among 20 runs of each benchmark. The shaded area indicates ±1 standard deviation.

The figure shows that most benchmarks have a throughput ratio close to 1. There are a few outliers with significantly lower throughput, nearly 10× worse than the original Ruby. Those outliers are benchmarks related to fiber operations. Fibers in Ruby are lightweight concurrency primitives that allow cooperative multitasking. The Ruby codebase contains several implementations for fiber to optimize based on the underlying architecture. In our ported Ruby, we adapted the general implementation that relies on standard library functions, while the original Ruby uses the implementation optimized for an x86 written in assembly.

On average, the ported Ruby runs at 0.982× throughput of the original Ruby, with a standard deviation of 0.038×.

## 6 Related Work

**Prior VM Ports to CHERI**    MicroPython is an implementation of Python to run on microcontrollers and in constrained environments. The porting of MicroPython [23] to CHERI describes the overall comprehensive process, including lessons from the challenges faced and the evaluation of the performance.

Harris et al. [17] discuss the challenges and solutions involved in porting the Rust programming language to CHERI. This work covers various aspects of the porting process, including modifications to the compiler and standard library. This case study also points out Rust-specific issues, such as the lack of counterpart to (u)intptr_t in Rust.

The porting of JavaScriptCore to CHERI [16] was also studied. This work evaluates the feasibility of porting a contemporary language runtime and the source code compatibility on CHERI.

**Undefined Behavior**    In implementing a memory-safe C for CHERI, Chisnall et al. [12] summarize common C idioms that originally caused errors in CHERI's memory model. A portion of them depend on implementation-defined or undefined behaviors. By supporting all but one of these idioms, the authors refined the CHERI ISA to enhance code compatibility. The excluded idiom refers to storing a pointer in a smaller size integer, and it is related to the issue we mentioned in Section 4.6. The rest of the originally problematic idioms are distinct to this work. This work investigated general open-source applications, while we focus on VM implementations.

Research on the semantics of undefined behavior includes Lenient C [25], a C dialect that assigns precise semantics to undefined, unspecified, and implementation-defined behaviors, improving portability. It is implemented in Safe Sulong, an interpreter with a dynamic compiler for the JVM. Hathhorn et al. [18] propose negative semantics for C, rejecting programs with undefined behaviors and providing a tool to detect them.

**Other System Software on CHERI**    Other than VMs, there are also works that port other kinds of system software to CHERI. The paper about CheriABI [13] proposes the concept of *abstract capability* to describe memory access. To investigate this concept, the authors ported FreeBSD and PostgreSQL to CHERI. The changes in the adaptation process are summarized and categorized into 11 categories. Their categories cover most of our issues, but we illustrated what issues arise in the context of VMs.

Jacob and Singer [20] describe the experience of adapting BDW conservative GC library [5] to CHERI in the motivation of securing the GC metadata. It is the first analysis of the complexities of porting GC to CHERI. They show the challenges in porting, performance evaluation, and opportunities for runtime optimizations. As in our porting, they use the validity tag to exclude non-capability values in the marking phase.

## 7 Conclusions

In this work, we ported CRuby to CHERI and surveyed three prior case studies to identify pitfalls in developing and porting VMs to CHERI. We classified these pitfalls into six categories and examined their workarounds and impacts. Most of the pitfalls arose from the undefined behaviors in the C language; in particular, implementation techniques and idioms of VMs often assume behaviors of traditional architectures that are invalid on CHERI. For example, conservative GC depends on that pointers and integers are interchangeable; bitmaps and bitfields, which embed small fields in an integer, assume that all bits of integer types are available, whereas (u)intptr_t is a capability on CHERI; and deriving a pointer from another created for different purposes is common in VMs.

We showed that these pitfalls can be addressed by maintaining a capability with wide bounds, checking validity tag in GC, avoiding relying on in-place reallocation, and using appropriate integer types. Although some workarounds introduce performance overhead or weaken CHERI's security guarantees, CHERI also enables more accurate GC.

**Acknowledgements** Parts of this work are supported by JSPS through JSPS KAKENHI grant number JP23K24822. We would also like to thank Jeremy Singer and Kai Feng of the University of Glasgow, the CHERI team, and the Ruby community for their support during this project.

**Pitfalls in VM Implementation on CHERI: Lessons from Porting CRuby**

## About the authors

**Hanhaotian Liu** hht@g.ecc.u-tokyo.ac.jp.
https://orcid.org/0009-0006-1223-6987

**Tetsuro Yamazaki** yamazaki@csg.ci.i.u-tokyo.ac.jp.
https://orcid.org/0000-0002-2065-5608

**Tomoharu Ugawa** tugawa@acm.org.
https://orcid.org/0000-0002-3849-8639